\documentclass[aip, jcp, reprint]{revtex4-1} 

\usepackage[english]{babel}
\usepackage[utf8x]{inputenc}
\usepackage{amsmath,amsfonts,amssymb,amsbsy,eucal,dsfont}
\usepackage{color}
\usepackage[colorlinks]{hyperref}
\usepackage{graphicx}
\graphicspath{{Figures/}}

\newcommand{\hide}[1]{}

\newcommand{\kT}{k_B T}

\newcommand{\ie}{\textit{i.e.}}
\newcommand{\eg}{\textit{e.g.}}
\newcommand{\etc}{\textit{etc.}}

\newcommand{\bF}{\ensuremath{\boldsymbol{F}}}

\newcommand{\bU}{\ensuremath{\boldsymbol{U}}}

\newcommand{\br}{\ensuremath{\boldsymbol{r}}}

\newcommand{\be}{\ensuremath{\boldsymbol{e}}}

\newcommand{\bS}{\ensuremath{\boldsymbol{S}}}

\newcommand{\bmu}{\ensuremath{\boldsymbol\mu}}

\newcommand{\bR}{\ensuremath{\boldsymbol{R}}}

\newcommand{\Sab}{\ensuremath{S_{\alpha\beta}}}

\begin{document}

\title{Short-time diffusion in concentrated bidisperse hard-sphere suspensions}

\author{Mu Wang}
\email[]{mwwang@caltech.edu}
\affiliation{Division of Chemistry and Chemical Engineering, California Institute of Technology, Pasadena,
California 91125, USA}

\author{Marco Heinen}
\email[]{mheinen@caltech.edu}
\affiliation{Division of Chemistry and Chemical Engineering, California Institute of Technology, Pasadena,
California 91125, USA}

\author{John F. Brady}
\email[]{jfbrady@caltech.edu}
\affiliation{Division of Chemistry and Chemical Engineering, California Institute of Technology, Pasadena,
California 91125, USA}

\date{\today}

\renewcommand{\figurename}{Fig.}
\renewcommand{\figuresname}{Figs.}
\renewcommand{\refname}{Ref.}
\newcommand{\refsname}{Refs.}   
\newcommand{\expressionname}{Eq.}
\newcommand{\expressionsname}{Eqs.}
\newcommand{\sectionname}{Sec.}
\newcommand{\sectionsname}{Secs.}

\begin{abstract}
Diffusion in bidisperse Brownian hard-sphere suspensions is studied by Stokesian Dynamics (SD) computer simulations and 
a semi-analytical theoretical scheme for colloidal short-time dynamics, based on Beenakker and Mazur's method [\textit{Physica} \textbf{120A}, 388 (1983) \& \textbf{126A}, 349 (1984)].
Two species of hard spheres are suspended in an overdamped viscous solvent that mediates the salient hydrodynamic interactions among all particles.
In a comprehensive parameter scan that covers various packing fractions and suspension compositions, we employ numerically accurate SD simulations to compute the initial diffusive relaxation of density modulations at the Brownian time scale, quantified by the partial hydrodynamic functions.
A revised version of Beenakker and Mazur's $\delta\gamma$-scheme for monodisperse suspensions is found to exhibit surprisingly good accuracy, when simple rescaling laws are invoked in its application to mixtures.
The so-modified $\delta\gamma$ scheme predicts hydrodynamic functions in very good agreement with our SD simulation results, for all densities from the very dilute limit up to packing fractions as high as $40\%$.
\end{abstract}

\pacs{82.70.Dd. 
      82.70.Kj, 
      66.10.cg, 
      }
\maketitle

\section{Introduction}\label{sec:Intro}
Short-time diffusion in Brownian suspensions has been a topic of extensive research for many years, which has pushed forward the development of various computer simulation methods including Lattice Boltzmann simulations \cite{Ladd1993,Lobaskin2004,Duenweg2009}, Dissipative Particle Dynamics \cite{Hoogerbrugge1992,Li2008}, Stochastic Rotation / Multiparticle Collision Dynamics \cite{Malevanets1999,Ihle2001,Wysocki2010}, hydrodynamic force multipole methods \cite{Ladd1990,Cichocki1999}, boundary integral methods \cite{Pozrikidis1992,Kumar2012}, and (Accelerated) Stokesian Dynamics \cite{Brady1988,Sierou2001,BanchioBrady2003}.
Each of these simulation methods is rather involved, which is one reason for the on-going development of approximate (semi-) analytical theoretical schemes for colloidal short-time dynamics \cite{Jones1979,Beenakker1983,BeenakkerMazur1983,Batchelor1983,Mazur1984,Nagele1996,Zhang2002,Banchio2008,Heinen2011dyn,Makuch2012}.

In spite of extensive simulations and analytical theoretical studies, substantial gaps remain in the colloidal suspension parameter space that has not yet been explored, which is due both to the large number of tunable parameters in soft matter systems, and the complexity of the salient hydrodynamic interactions (HIs) among the suspended particles. 
The purpose of the present work is to assess the short-time diffusive dynamics in \textit{mixtures} of hard spheres with two different hard-core diameters using a generalization of Stokesian Dynamics (SD) simulations and an analytical-theoretical scheme.
While similar studies have so far been limited to suspensions in which at least one of the species is very dilute \cite{Jones1979,Batchelor1983,Zhang2002,Krueger2009}, in the present article we cover a large range of packing fractions including both dilute and dense bidisperse hard-sphere fluids. 
All results presented here can be straightforwardly generalized to suspensions of more than two particle species.

In addition to the steric no-overlap constraint, the suspended hard spheres interact via solvent-mediated HIs. 
Accurate inclusion of HIs into theory and simulation is essential, since the linear transport coefficients for colloidal suspensions are governed entirely by the HIs in the colloidal short-time regime. 
However, the peculiar properties of HIs render their computation a formidable task. 
In particular, HIs are long-ranged, non-pairwise-additive, and exhibit steep divergences in case of lubrication, \textit{i.e.}, when particles move in close contact configurations.

A semi-analytical theoretical scheme for short-time suspension dynamics, with multi-body HIs included in an approximate fashion, has been devised by Beenakker and Mazur \cite{Beenakker1983,BeenakkerMazur1983,Mazur1984}, and has quite recently been re-assessed by Makuch and Cichocki \cite{Makuch2012}. 
This method, commonly referred to as the $\delta\gamma$ scheme, makes use of resummation techniques by which an infinite subset of the hydrodynamic scattering series \cite{Makuch2012_scatteringseries} is computed, including all particles in suspension. 
Nevertheless, a complementary infinite subset of scattering diagrams is omitted in the $\delta\gamma$ scheme which, moreover, fails to include the correct lubrication limits of particle mobilities. 
Comparisons of the original $\delta\gamma$-scheme predictions to experimental and computer simulation data have revealed a shortcoming of the $\delta\gamma$ scheme in its prediction of self-diffusion coefficients \cite{Genz1991, Banchio2008, Heinen2011dyn, Westermeier2012, Makuch_in_prep}, which can be largely overcome by resorting to a modified $\delta\gamma$ scheme in which the computation of the self-diffusion coefficients is carried out by a more accurate method \cite{Banchio2008, Heinen2011dyn, Westermeier2012}.
To date the (modified) $\delta\gamma$ scheme remains the only analytical-theoretical approach that captures the essential physics of diffusion in dense suspensions, making predictions at an acceptable accuracy level. 
Unfortunately, the $\delta\gamma$ scheme has so far been formulated for monodisperse suspensions only, and a stringent generalization to mixtures poses a tedious task. 

Here we propose a simple rescaling rule that allows the application of the numerically efficient, easy to implement standard $\delta\gamma$-scheme expressions to mixtures of bidisperse hard spheres. 
The rescaling rule is based on the notion of describing either species as an effective, structureless host medium for the other species to move in. 
By comparing to our SD simulation results we show that the rescaled, modified $\delta\gamma$ scheme predicts both species' partial hydrodynamic functions with a surprisingly good accuracy, for suspension volume fractions as high as $40\%$.
The proposed, rescaled $\delta\gamma$ scheme can be particularly useful in the analysis of scattering experiments, where only a limited part of the hydrodynamic function can be measured due to the limited range of accessible wave vectors.

The remaining part of this article is organized as follows: 
In \sectionname~\ref{sec:HS_mixtures} we define the hard-sphere mixtures under study, and discuss the prevailing interactions among the particles. 
Section~\ref{sec:Shorttime_diff} contains a discussion of colloidal short-time diffusion and the partial hydrodynamic functions that are calculated in the present work.
Our SD simulations are outlined in \sectionname~\ref{sec:SD}, which is followed by a description of the rescaled $\delta\gamma$ scheme in \sectionname~\ref{sec:dg}.
In \sectionname~\ref{sec:Results} we present our results for partial hydrodynamic functions of various suspensions, and we draw our finalizing conclusions in \sectionname~\ref{sec:Conclus}.

\section{Bidisperse hard-sphere suspensions}\label{sec:HS_mixtures}
We study unbounded homogeneous equilibrium suspensions of non-overlapping Brownian hard spheres with hard-core radii $a_\alpha$ and $a_\beta$.
The pairwise additive direct interaction potentials between the particles can be written as
\begin{equation}\label{eq:pair_pot}
u_{\alpha\beta}(r) = \left\lbrace
   \begin{array}{ll}
   \infty\,& ~~\text{for}~r < a_\alpha + a_\beta,\\~\\
   0\,& ~~\text{otherwise}\\
   \end{array}
 \right. \\
\end{equation}
in terms of the particle-center separation distance $r$ and the particle species indices $\alpha, \beta \in \left\lbrace 1,2
\right\rbrace$. 
The suspensions' thermodynamic equilibrium state, studied in the present work, is entirely described by the three non-negative dimensionless parameters
\begin{eqnarray}
\lambda &=& a_2 / a_1, \label{eq:size_ratio}\\
~\nonumber\\
\phi &=& \phi_1 + \phi_2, ~~\text{and} \label{eq:total_volfrac}\\
~\nonumber\\
y &=& \phi_1 / \phi, \label{eq:volume_fraction_fraction}
\end{eqnarray}
where $\lambda$ is the size ratio and $\phi_\alpha = (4/3) \pi n_\alpha a_\alpha^3$ is the volume fraction of species $\alpha$ in terms of the partial number concentration $n_\alpha = N_\alpha / V$. 
In taking the thermodynamic limit both the number, $N_\alpha$, of particles of species $\alpha$, and the system volume $V$ diverge to infinity while their ratio $n_\alpha$ is held fixed. 
The remaining parameters in \expressionsname~\eqref{eq:total_volfrac} and \eqref{eq:volume_fraction_fraction} are the total volume fraction $\phi$ and the composition ratio $y$, which satisfies $0 \leq y \leq 1$.
Without loss of generality, we assume $a_2 \geq a_1$ in the following.
We denote the total number of particles as $N$, and obviously, $N = N_1 + N_2$.

All particles are assumed neutrally buoyant in an infinite quiescent, structureless Newtonian solvent of shear viscosity $\eta_0$. 
No external forces or torques act on the suspended particles. 
The solvent is assumed to be incompressible, and the Reynolds number for particle motion is assumed to be very small, such that the solvent velocity field $\boldsymbol{v}(\boldsymbol{r})$ and  dynamic pressure field $p(\boldsymbol{r})$ satisfy the stationary Stokes equation with incompressibility constraint, 
\begin{eqnarray}
\eta_0 \Delta \boldsymbol{v}(\boldsymbol{r}) &=& \nabla p(\boldsymbol{r}), \label{eq:Stokes_eqn}\\
~\nonumber\\
\nabla \cdot \boldsymbol{v}(\boldsymbol{r}) &=& 0 \label{eq:incompressibility},
\end{eqnarray}
at every point $\boldsymbol{r}$ inside the solvent.
Equations~\eqref{eq:Stokes_eqn} and \eqref{eq:incompressibility} are supplemented with hydrodynamic no-slip boundary conditions on the surface of each suspended sphere.
The linearity of Eq.~(\ref{eq:Stokes_eqn}) and (\ref{eq:incompressibility}) suggests a linear coupling between the translational velocity of particle $l$, $\bU_l$, and the force exerted on particle $j$, $\bF_j$:
\begin{equation}
  \bU_l = -\bmu_{lj}^{tt} \cdot \bF_j,
\end{equation}
where the mobility tensor $\bmu_{lj}^{tt}$ has a size of $3 \times 3$. 
By placing the tensor $\bmu_{ij}^{tt}$ as elements of a larger, generalized matrix, we construct the suspension grand mobility tensor $\bmu^{tt}$ of size $3N \times 3N$.
The minimum dissipation theorem~\cite{KimKarilla} requires $\bmu^{tt}$ to be symmetric and positive definite.


\section{Short-time diffusion}\label{sec:Shorttime_diff}
Here we are interested in diffusive dynamics at a coarse-grained scale of times $t$ that satisfy the two strong inequalities~\cite{Dhont1996}
\begin{equation}\label{eq:short-time-regime}
\tau_H \sim \tau_I \ll t \ll \tau_D,
\end{equation}
defining the colloidal \textit{short-time} regime.
The hydrodynamic time scale $\tau_H = a_2 \rho_0 / \eta_0$, involving the solvent mass density $\rho_0$, quantifies the time at which solvent shear waves traverse typical distances between (the larger) colloidal particles. 
The criterion $t \gg \tau_H$ implies that HIs, being transmitted by solvent shear waves, act effectively instantaneously at the short-time scale.
Therefore, the elements of the grand mobility matrix $\bmu^{tt}$ depend on the instantaneous positions $\boldsymbol{r}^N=\{\br_1, \br_2, \cdots, \br_N\}$ of all particles, but not on their positions at earlier times.
The momentum relaxation time $\tau_I = m_2 / (6 \pi \eta_0 a_2)$ in terms of the mass, $m_2$ of a particle of species 2, is similar in magnitude to $\tau_H$. 
At times $t \gg \tau_I$, many random collisions of a colloidal particle with solvent molecules have taken place, the particle motion is diffusive, and the inertia plays no role. 
The colloidal short time regime is bound from above by the (diffusive) interaction time scale $\tau_D = a_1^2 / d_0^1$, given in terms of the Stokes-Einstein-Sutherland (SSE) translational free diffusion coefficient, $d_0^1 = k_B T\mu_0^1$ of the smaller particle species. 
Here, $\mu_0^\alpha=(6\pi\eta_0 a_\alpha)^{-1}$ is the single particle mobility of species $\alpha$, $k_B$ is the Boltzmann constant and $T$ is the absolute temperature. 
During times $t \gtrsim \tau_D$, diffusion causes the spatial configuration of the (smaller) particles to deviate appreciably from their initial configuration, and in addition to the HIs, rearrangements of the cage of neighboring particles start to influence particle dynamics. 
This results in a sub-diffusive particle motion at times $t \gtrsim \tau_D$ preceding the ultimate diffusive long-time regime $t \gg \tau_D$ at which a particle samples many independent local neighborhoods.
Unless the particle size-ratio $\lambda$ is very large, $\tau_D$ is some orders of magnitude larger than both $\tau_H$ and $\tau_I$, and the colloidal short-time regime in \expressionname~\eqref{eq:short-time-regime} is well defined \cite{Dhont1996}.

Scattering experiments on bidisperse colloidal suspensions, including the most common small angle light scattering \cite{BernePecora1976} and x-ray scattering \cite{Guinier1955,Kratky1982} techniques, allow the extraction of the measurable dynamic structure factor
\cite{Nagele1996} 
\begin{equation}\label{eq:S_measurable}
S_M(q,t) = \dfrac{1}{\overline{f^2}(q)} \sum\limits_{\alpha,\beta=1}^2 \sqrt{x_\alpha x_\beta} ~f_\alpha(q) f_\beta(q) ~S_{\alpha\beta}(q,t),
\end{equation}
which contains the scattering amplitudes, $f_\alpha(q)$, for particles of either species, the mean squared scattering amplitude $\overline{f^2}(q) = x_1 f_1^2(q) + x_2 f_2^2(q)$ in terms of the molar fractions $x_\alpha = N_\alpha/N$, and the partial dynamic structure factors $S_{\alpha\beta}(q, t)$. 
In case of scattering experiments, $N_\alpha$ is the mean number of $\alpha$-type particles in the scattering volume.
The microscopic definition of the partial dynamic structure factors reads
\begin{equation}\label{eq:S_ab_microscop_definition}
\hspace{-.6em}
S_{\alpha\beta}(q,t) = \lim\limits_\infty \left\langle \frac{1}{\sqrt{N_\alpha N_\beta}}
\sum_{l=l_{\text{min}}\atop j=j_{\text{min}}}^{l_{\text{max}}\atop j_{\text{max}}}
\exp\left\lbrace {i \mathbf{q}\cdot[\br_l^{\alpha}(0) - \br_j^{\beta}(t)]} \right\rbrace \right\rangle ,
\end{equation}
where $l_{\text{min}} = (\alpha-1)N_\beta + 1$ and $j_{\text{min}} = (\beta-1)N_\alpha + 1$ are the lower summation range limits,
$l_{\text{max}} = l_{\text{min}} + N_\alpha - 1$ and $j_{\text{max}} = j_{\text{min}} + N_\beta - 1$ are the upper summation range limits,
$i = \sqrt{-1}$ is the imaginary unit, $\lim_\infty$ indicates the thermodynamic limit, the brackets $\left\langle \ldots \right\rangle$ stand for the ensemble average, and $\br_k^{\gamma}(t)$ is the position of particle number $k$ (which belongs to species $\gamma$) at time $t$. 
From the microscopic definition it follows that $S_{\alpha\beta}(q) = S_{\beta\alpha}(q)$, and that the functions $S_{\alpha\alpha}(q)$ are non-negative, while the $S_{\alpha\beta}(q)$ for $\alpha \neq \beta$ can assume either sign. 
In the special case of $t=0$, the partial dynamic structure factors reduce to the partial static structure factors $S_{\alpha\beta}(q) = S_{\alpha\beta}(q, 0)$ and, likewise, $S_M(q,0) = S_M(q)$ is the measurable static structure factor. 

A useful approximation in the analysis of experimental scattering data for suspensions with a small degree of particle polydispersity (typically $10\%$ or less relative standard deviation in the particle-size distribution) is the decoupling approximation \cite{Nagele1996, Westermeier2012} in which all functions $S_{\alpha\beta}(q,t)$ in \expressionname~\eqref{eq:S_measurable} are approximated by a monodisperse, mean structure factor $S(q,t)$. 
For the strongly size-asymmetric hard-sphere mixtures studied here, the $S_{\alpha\beta}(q,t)$ show distinct mutual differences, which rules out the application of the decoupling approximation.

In some experiments, the $f_\alpha(q)$ for different species $\alpha$ may be tuned independently. 
An example is the selective refractive index matching of solvent and particles in light scattering experiments \cite{Williams2001}. 
Under such circumstances, the three independent functions $S_{\alpha\beta}(q)$ for $\alpha,\beta \in \left\lbrace 1, 2 \right\rbrace$ may be singled out individually. 
When all functions  $S_{\alpha\beta}(q,t)$ are known, the dynamic number-number structure factor 
\begin{equation}\label{eq:S_NN}
S_{NN}(q,t) = \sum\limits_{\alpha,\beta=1}^2 \sqrt{x_\alpha x_\beta} ~S_{\alpha\beta}(q,t),
\end{equation}
can be determined which reduces, for $t=0$, to the static number-number structure factor $S_{NN}(q)$. 
In computer simulations, the $S_{\alpha\beta}(q,t)$ and $S_{NN}(q,t)$ are easily extracted once that all the time-dependent particle positions $\mathbf{r}_k^{\gamma}(t)$ are known, but the challenge lies in the accurate computation of the latter.

Colloidal dynamics at times $t \gg \tau_H \sim \tau_B$ are governed by the Smoluchowski diffusion equation \cite{Dhont1996} which quantifies the temporal evolution for the probability density function $P(t, \br^N)$ of the particle configuration $\br^N$ at time $t$. 
It can be shown \cite{Akcasu1984} that the $2 \times 2$ correlation matrix $\boldsymbol{S}(q,t)$ with elements $S_{\alpha\beta}(q,t)$ decays at short times as 
\begin{equation}\label{eq:Sqt_exp_decay}
\boldsymbol{S}(q,t) = e^{-q^2 \boldsymbol{D}(q) t} \cdot \boldsymbol{S}(q), 
\end{equation}
with a diffusivity matrix $\boldsymbol{D}(q)$ that can be split as
\begin{equation}\label{eq:Dq_matrix}
\boldsymbol{D}(q) = \kT\boldsymbol{H}(q) \cdot \boldsymbol{S}^{-1}(q), 
\end{equation}
into a product of the matrix $\boldsymbol{H}(q)$ of partial hydrodynamic functions $H_{\alpha\beta}(q)$ and the inverse partial static structure factor matrix $\boldsymbol{S}^{-1}(q)$.

The functions $H_{\alpha\beta}(q)$ can be interpreted as generalized wavenumber-dependent short-time sedimentation velocities: 
In a homogeneous suspension, the value of $H_{\alpha\beta}(q)$ quantifies the spatial Fourier components of the initial velocity attained by particles of species $\alpha$, when a weak force field is switched on that acts on particles of species $\beta$ only, dragging them in a direction parallel to $\mathbf{q}$ with a magnitude that oscillates harmonically as $\cos(\mathbf{q} \cdot \br)$.
The microscopic definition of the partial hydrodynamic functions reads \cite{Nagele1996} 
\begin{widetext}
\begin{equation}\label{eq:H_ab_microscop_definition}
H_{\alpha\beta}(q) = \lim\limits_\infty \left\langle \frac{1}{\sqrt{N_\alpha N_\beta}}
\sum_{l=l_{\text{min}}\atop j=j_{\text{min}}}^{l_{\text{max}}\atop j_{\text{max}}}
\hat{\mathbf{q}} \cdot {\boldsymbol{\mu}^{tt}_{lj}(\br^N)}
       \cdot \hat{\mathbf{q}} ~\exp\left\lbrace {i \mathbf{q}\cdot[\br_l^{\alpha} - \br_j^{\beta}]} \right\rbrace \right\rangle ,
\end{equation}
\end{widetext}
where $\hat{\mathbf{q}} = \mathbf{q} / q$ is the normalized wave vector, and the summation ranges $l_{\text{min}} \ldots l_{\text{max}}$ and $j_{\text{min}} \ldots j_{\text{max}}$ are the same as \expressionname~\eqref{eq:S_ab_microscop_definition}.
Note that the positive definiteness of the $\boldsymbol{\mu}^{tt}$ implies that the functions $H_{\alpha\alpha}(q)$ are non-negative, whereas the functions $H_{\alpha\beta}(q)$ can assume both positive and negative values for $\alpha \neq \beta$. 
In particular, the latter functions assume negative values at small values of $q$ due to the solvent backflow effect: 
When a weak spatially homogeneous external force acts on particles of species $\beta$ only, it causes the $\beta$-type particles to sediment in a direction parallel to the applied force, which corresponds to $H_{\beta\beta}(q \to 0) > 0$. 
Mass conservation requires the collective motion of $\beta$-type particles to be compensated by an opposing backflow of solvent, which drags the $\alpha$-type particles in the direction anti-parallel to the applied force. 
Hence, $H_{\alpha\beta}(q \to 0) < 0$ for $\alpha \neq \beta$. 

By splitting the sum in \expressionname~\eqref{eq:H_ab_microscop_definition} into the self ($l=j$) and the complementary distinct contributions, the functions $H_{\alpha\beta}(q)$ can each be decomposed, according to
\begin{equation}\label{eq:Hq_self_distinct_splitting}
H_{\alpha\beta}(q) = \delta_{\alpha\beta} \dfrac{d_s^{\alpha}}{k_B T} + H_{\alpha\beta}^d(q),
\end{equation}
into a sum of a wavenumber-independent self-part and the wavenumber-dependent distinct part of the partial hydrodynamic function, $H_{\alpha\beta}^d(q)$, which tends to zero for large values of $q$. 
In case of infinite dilution, or in the (purely hypothetical) case of vanishing hydrodynamic forces, $H_{\alpha\beta}(q)/\mu_0^\alpha$ reduces to the Kronecker delta symbol $\delta_{\alpha\beta}$. 
The short-time translational self diffusion coefficient ${d_s^{\alpha}}$ is equal to the time derivative of the mean squared
displacement $W_\alpha(t) = \tfrac{1}{6} \left\langle   {[\boldsymbol{r}_l^\alpha(t) - \boldsymbol{r}_l^\alpha(0)]}^2 \right\rangle$ of a particle of species $\alpha$ at short times. 
At infinite dilution, ${d_s^{\alpha}} = {d_0^{\alpha}}$. 

If all functions $H_{\alpha\beta}(q)$ are known, then the number-number hydrodynamic function
\begin{equation}\label{eq:H_NN}
H_{NN}(q) = \sum\limits_{\alpha,\beta=1}^2 \sqrt{x_\alpha x_\beta} ~H_{\alpha\beta}(q)
\end{equation}
and the measurable hydrodynamic function
\begin{equation}\label{eq:H_M}
H_M(q,t) = \dfrac{1}{\overline{f^2}(q)} \sum\limits_{\alpha,\beta=1}^2 \sqrt{x_\alpha x_\beta} ~f_\alpha(q) f_\beta(q) ~H_{\alpha\beta}(q),
\end{equation}
can be computed, which quantify the short-time decay of the dynamic number-number structure factor 
\begin{equation}\label{eq:S_NN_exp_decay}
S_{NN}(q,t) = S_{NN}(q) e^{-q^2 D_{NN}(q) t}
\end{equation}
and the measurable dynamic structure factor
\begin{equation}\label{eq:S_M_exp_decay}
S_M(q,t) = S_M(q) e^{-q^2 D_M(q) t},
\end{equation}
through the number-number diffusion function $D_{NN}(q) = k_B T H_{NN}(q) / S_{NN}(q)$ and the measurable diffusion function $D_M(q) = k_B T H_M(q) / S_M(q)$.

\section{Stokesian Dynamics simulations}\label{sec:SD}


The framework of the Stokesian Dynamics (SD) has been extensively discussed elsewhere~\cite{sd_durlofsky_jfm1987,brady-sd-ew_jfm_88,Brady1988,BanchioBrady2003} and here we only present the aspects pertinent to the present work. 
For rigid particles in a suspension, the generalized particle forces $\boldsymbol{\cal F}$ and stresslets $\bS$ are linearly related to the generalized particle velocities $\boldsymbol{\cal U}$ through the grand resistance tensor $\boldsymbol{\cal R}$ as~\cite{KimKarilla} 
\begin{equation}
\begin{pmatrix}\boldsymbol{\cal F} \\ \bS \end{pmatrix} = 
- \boldsymbol{\cal R}\cdot
\begin{pmatrix}\boldsymbol{\cal U} - \boldsymbol{\cal U}^\infty\\-\be^\infty\end{pmatrix},  
\end{equation}
where $\boldsymbol{\cal U}^\infty$ and $\be^\infty$ are the imposed generalized velocity and strain rate, respectively.
The generalized force $\boldsymbol{\cal F}$ represents the forces and torques of all particles in the suspension, and the generalized velocity $\boldsymbol{\cal U}$ contains the linear and angular velocities for all particles.
The grand resistance tensor $\boldsymbol{\cal R}$ is partitioned as 
\begin{equation}
\boldsymbol{\cal R} = \begin{pmatrix}\bR_{FU} & \bR_{FE} \\ \bR_{SU} & \bR_{SE} \end{pmatrix},  
\end{equation}
where, for example, $\bR_{FU}$ describes the coupling between the generalized force and the generalized velocity, $\bR_{FE}$ describes the coupling between the generalized force and the strain rate, \etc.
In the SD method the grand resistance is approximated as 
\begin{equation}
\boldsymbol{\cal R} = (\boldsymbol{\cal M}^{\infty})^{-1} + \boldsymbol{\cal R}_{2B} - \boldsymbol{\cal R}_{2B}^\infty,
\end{equation}
where the far field mobility tensor $\boldsymbol{\cal M}^{\infty}$ is constructed pairwisely from the multipole expansions and the Fax\'{e}n's laws of the Stokes equation up to the stresslet level, and its inversion captures the long-range many-body hydrodynamic interactions.
The near-field lubrication correction $(\boldsymbol{\cal R}_{2B} -\boldsymbol{\cal R}_{2B}^\infty)$ is based on the exact two-body solutions with the far field contributions removed, and it accounts for the singular HIs when particles are in close contact.
The SD method recovers the exact solutions of the two-particle problems and was shown to agree well with the exact solution of three-particle problems~\cite{three-spheres_wilson_jcp2013}.

\begin{figure}[ht]
  \centering
    \includegraphics[width=3in]{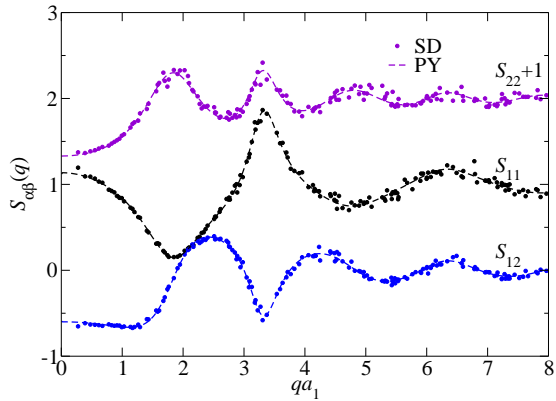}
  \caption{(Color online) The bidisperse suspension partial static structure factors $S_{\alpha\beta}(q)$ directly measured from the simulations (dots) and computed via the Percus-Yevick integral equation scheme (dashed curves) for $\phi =   0.5$, $y = 0.5$, and $\lambda = 2$.  
Note that $S_{22}(q)$ has been shifted up by 1 for clarity.  \label{fig:sk}} 
\end{figure}

Extending the SD method to polydisperse systems retains the computational framework above.
The far-field polydisperse mobility tensor $\boldsymbol{\cal M}^{\infty}$ is computed using the multipole expansions as in \refsname~\onlinecite{sd-bimodal-monolayer_powell_jfm1993} and the resulting expressions are extended to infinite periodic systems using Beenakker's technique~\cite{ewald-sum-rotne-prager_beenaker_jcp86,ewald-poly_hase_pof01}.
The lubrication correction $(\boldsymbol{\cal R}_{2B}-\boldsymbol{\cal   R}_{2B}^\infty)$ for particle pair with radii $a_\alpha$ and $a_\beta$ are based on the exact solution of two-body problems in \refsname~\onlinecite{Jeffrey1984, resist-func_jeffrey_pof1992, pres-moment_jeffrey_pof1993, compres-res_khair_pof2006} up to $s^{-300}$, where $s = 2r/(a_\alpha+a_\beta)$ is the scaled center-to-center particle distance.
In our simulations, the lubrication corrections are invoked when $r<2(a_\alpha + a_\beta)$, and the analytic lubrication expressions are used when $r<1.05(a_\alpha + a_\beta)$.


Our simulations proceed as follows.  First, a random bimodal hard-sphere packing at the desired composition is generated using the event-driven Lubachevsky-Stillinger algorithm~\cite{Lubachevsky1990,packing-gen-code_torquato_pre2006} with high compression rate. 
After the desired volume fraction $\phi$ is reached, the system is equilibrated for a short time (10 events per particle) at zero compression rate.  
This short equilibration stage is necessary as the compression pushes particles closer to each other. Prolonging the equilibration stage does not alter the resulting suspension structure significantly.

Figure~\ref{fig:sk} features the partial static structure factors $S_{\alpha\beta}(q)$ from the above simulation protocol for a bidisperse suspension of $\lambda = 2$, $y=0.5$, and $\phi = 0.5$, the highest volume fraction studied in this paper.
The measured functions $S_{\alpha\beta}(q)$ are compared with the Percus-Yevick (PY)\cite{Percus1958, sk-binary-mix_ashcroft_pr1967,  sk-binary-mix-errata_ashcroft_pr1968}
integral equation solutions at the same composition. We use the analytical PY scheme solutions as input to the rescaled $\delta\gamma$ scheme, as described in the following section.   
Note that even at the high volume fraction $\phi = 0.5$, the PY closure describes the suspension structure very accurately. 

To avoid singularities in the grand resistance tensor due to particle contact, we enforce a minimum separation of $10^{-6}(a_i+a_j)$ between particles in our simulations.
The resistance tensor $\boldsymbol{\cal R}$ is then constructed based on the particle configuration $\br^N$.
The partial hydrodynamic functions are extracted from $\bmu^{tt}$, a submatrix of the grand mobility tensor
\begin{equation}
  \bR_{FU}^{-1} = 
\begin{pmatrix}
    \bmu^{tt} & \bmu^{tr} \\
    \bmu^{rt} & \bmu^{rr} \\
\end{pmatrix},
\end{equation}
which contains coupling between the translational (t) and rotational (r) velocities and forces of a freely-mobile particle suspension. 
Typically each configuration contains $800$ particles and at least $500$ independent configurations are studied for each composition.



The partial hydrodynamic functions $H_{\alpha\beta}(q)$ extracted from the simulations exhibit a strong $\sqrt[3]{N}$ size dependence due to the imposed periodic boundary conditions~\cite{sd-ew-transport-coeff-pt1_phillips_pof1988, Ladd1990, Ladd:95,Banchio2008}.  
The finite size effect can be eliminated by considering $H_{\alpha\beta}(q)$ as a generalized sedimentation velocity.
The sedimentation velocity from a finite size system with periodic boundary conditions is a superposition of the velocities from random suspensions and cubic lattices~\cite{sd-ew-transport-coeff-pt1_phillips_pof1988, Ladd:95}.
This argument is straightforwardly extended to bidisperse suspensions, where
the size correction, $\Delta_N H_{\alpha\beta}(q)$, for the partial hydrodynamic functions computed from the $N$-particles system, $H_{\alpha\beta,N}(q)$,
is 
\begin{equation}
  \label{eq:size-corr}
\Delta_N H_{\alpha\beta}(q) =  \frac{1.76 \mu_{0}^1
  [1+(\lambda^3-1)y]^{\frac{1}{3}} S_{\alpha\beta}(q) }{\lambda}
\frac{\eta_s}{\eta_0} \left( \frac{\phi}{N} \right)^{\frac{1}{3}}.  
\end{equation}
In \expressionname~\eqref{eq:size-corr}, $\Delta_N H_{\alpha\beta}(q) = H_{\alpha\beta}(q)-H_{\alpha\beta,N}(q)$, $H_{\alpha\beta}(q)$ is the hydrodynamic function in the thermodynamic limit, and $\eta_s/\eta_0$ is the high frequency shear viscosity of the suspension, which is obtained from the same simulation.
Note that the shear viscosity $\eta_s/\eta_0$ changes little with system size, and that the scaling for $H_{\alpha\beta}(q)$ in Eq.~(\ref{eq:size-corr}) is chosen to be $\mu_0^1$ regardless of the choice of $\alpha$ and $\beta$.

\section{Rescaled $\delta\gamma$ scheme}\label{sec:dg}

The $\delta\gamma$ scheme, originally introduced by Beenakker and Mazur \cite{BeenakkerMazur1983, Mazur1984} and quite recently revised by Makuch \textit{et al.} \cite{Makuch2012, Makuch_in_prep} predicts short-time linear transport coefficients of monodisperse colloidal suspensions with an overall good accuracy, for volume fractions of typically less than $40\%$.
A modified version of the $\delta\gamma$ scheme with an improved accuracy has been proposed in \refsname~\onlinecite{Genz1991, Banchio2008,  Heinen2011dyn, Westermeier2012}. 
The modification consists of replacing the rather inaccurate, microstructure-independent $\delta\gamma$-scheme expression for the self-diffusion coefficient $d_s$ by a more accurate expression.
The hydrodynamic function for a monodisperse suspension is then calculated as the sum of this more accurate self-term and the distinct part of the hydrodynamic function, the latter retained from the original $\delta\gamma$ scheme (\textit{c.f.}, the special case of \expressionname~\eqref{eq:Hq_self_distinct_splitting} for monodisperse suspensions). 
This replacement of the self-diffusion coefficient does not only result in an improved accuracy of the predicted hydrodynamic functions for hard spheres, but also allows computation of hydrodynamic functions of charge-stabilized colloidal particles with mutual electrostatic repulsion of variable strength. 

There are several possibilities for choosing the self-diffusion coefficient in the modified $\delta\gamma$ scheme. It can be treated as a fitting parameter \cite{Genz1991}, calculated by computer simulation \cite{Banchio2008}, or in the approximation of pairwise additive HIs, which is specially well-suited for charge-stabilized suspensions \cite{Heinen2011dyn, Westermeier2012}. 
In case of monodisperse hard-sphere suspensions,
\begin{equation}\label{eq:ds_HS_monodisp_virial}
\frac{d_s}{d_0} \approx 1 - 1.8315\phi (1 + 0.1195\phi - 0.70\phi^2),
\end{equation}
where $d_0= \kT \mu_0$ and $\mu_0 = (6\pi\eta_0 a)^{-1}$, is a highly accurate approximation provided that $\phi \lesssim 0.5$ \cite{Heinen2011dyn}.
Expression \eqref{eq:ds_HS_monodisp_virial} coincides with the known truncated virial expression \cite{Cichocki1999} to quadratic order in $\phi$. 
The prefactor of the cubic term has been determined as an optimal fit value that reproduces numerically precise computer simulation results for $d_s / d_0$ \cite{Banchio2008, Abade2011}.

The distinct part of the monodisperse hydrodynamic function is approximated in the $\delta\gamma$-scheme as:
\begin{eqnarray}\label{eq:Hdistinct_deltagamma}
\hspace{-.6em}
\frac{H^d(q)}{\mu_0} = \dfrac{3 }{2\pi} &&\int\limits_0^\infty dy' {\left[\dfrac{\sin(y')}{y'}\right]}^2 \cdot
{\left[1 + \phi S_{\gamma_0}(\phi, y')\right]}^{-1}\nonumber\\
\times &&\int\limits_{-1}^1 d\mu (1-\mu^2)\left[S(|\mathbf{q}-\mathbf{q'}|)-1\right].  
\end{eqnarray}
In \expressionname~\eqref{eq:Hdistinct_deltagamma}, $y = 2qa$ is a dimensionless wavenumber, $\mu = \mathbf{q} \cdot \mathbf{q'} /(q q')$ is the cosine of the angle between $\mathbf{q}$ and $\mathbf{q'}$, and the volume-fraction and wavenumber-dependent function $S_{\gamma_0}(\phi, y)$ (not to be confused with a static structure factor) has been specified in \refsname~\cite{Mazur1984, Genz1991}.

For monodisperse suspensions, the $\delta\gamma$ scheme requires only the static structure factor $S(q)$ and the suspension volume fraction $\phi$ as the input for calculating the hydrodynamic functions, namely,
\begin{equation}
  \frac{H(q)}{\mu_0} \approx H_{\delta\gamma}[ S(q), \phi],
\end{equation}
where $H_{\delta\gamma}[\cdot, \cdot]$ denotes the modified $\delta\gamma$-scheme result based on
\expressionsname~\eqref{eq:Hq_self_distinct_splitting}, \eqref{eq:ds_HS_monodisp_virial} and \eqref{eq:Hdistinct_deltagamma}.

Extending the $\delta\gamma$ scheme to the more general case of bidisperse suspensions is a non-trivial task.
The size polydispersity affects ($i$) the structural input through the partial static structure factors $\Sab(q)$, and ($ii$) the hydrodynamic scattering series \cite{Makuch2012_scatteringseries}, upon which the $\delta\gamma$ scheme is constructed \cite{Makuch2012}. 
For bidisperse suspensions, the structural input in ($i$) can be computed by liquid integral equations, \eg, the PY scheme~\cite{Percus1958, Lebowitz1964, sk-binary-mix_ashcroft_pr1967,  sk-binary-mix-errata_ashcroft_pr1968} which we use in the present study.  
However, the evaluation of the bidisperse hydrodynamic scattering series is more difficult since each scattering diagram for monodisperse suspensions has to be replaced by multiple diagrams describing the scattering in particle clusters containing particles of both species. 
Even if the resummation of the bidisperse hydrodynamic scattering series can be achieved, the accuracy of the results remains unknown without a direct comparison to experiments or computer simulations.

Here we bypass the difficult task of bidisperse hydrodynamic scattering series resummation and adopt a simpler idea based on the existing (modified) $\delta\gamma$ scheme for monodisperse particle suspensions.
The partial hydrodynamic functions $H_{\alpha\alpha}(q)$ can always be written as
\begin{equation}
\label{eq:hqaa-scal}
  \frac{H_{\alpha\alpha}(q)}{\mu_0^\alpha} = f_{\alpha} 
  H_{\delta\gamma}[S_{\alpha\alpha}(q), \phi_\alpha], 
\end{equation}
where the factor 
\begin{equation}
f_\alpha = f_{\alpha}(q; \lambda, \phi, y)
\end{equation}
describes the wave-number dependent HIs due to the other species $\beta$ not captured in the $\delta\gamma$ scheme, and also depends on the suspension composition.

For the interspecies partial hydrodynamic functions $H_{\alpha\beta}(q)$ ($\alpha\neq \beta$), the limiting value at $q\rightarrow \infty$, like $S_{\alpha\beta}(q)$, goes to zero.
Therefore, only the distinct part in the $\delta\gamma$ scheme is relevant, and to maintain consistency with Eq.~(\ref{eq:Hdistinct_deltagamma}), a shifted distinct static structure factor $S_{\alpha\beta}(q) + 1$ ($\alpha\neq\beta$) is used as the input.
Similar to Eq.~(\ref{eq:hqaa-scal}), a scaling factor $f_{\alpha\beta} = f_{\alpha\beta} (q; \lambda, \phi, y)$ provides the connection to the $\delta\gamma$ scheme by  
\begin{equation}
\label{eq:hqab-scal}
  \frac{H_{\alpha\beta}(q)}{ \mu_0^\alpha} = f_{\alpha\beta}  
  H_{\delta\gamma}^d [S_{\alpha\beta}(q)+1, \phi], (\alpha\neq\beta),
\end{equation}
when $H_{\delta\gamma}^d [S_{\alpha\beta}(q)+1, \phi]$ is computed according to \expressionname~\eqref{eq:Hdistinct_deltagamma}.
Note that in Eq.~(\ref{eq:hqab-scal}) the total volume fraction $\phi$ is used in the $\delta \gamma$ scheme.  
This is motivated by the physics of $H_{\alpha\beta}(q)$ ($\alpha\neq\beta$)---from a generalized sedimentation perspective, it describes the $q$-dependent velocity response of species $\alpha$ due to an application of $q$-dependent forces on the $\beta$ species.
Since both species are present, the total volume fraction $\phi$ should be used.
For monodisperse suspensions with artificially labeled particles, we expect $f_{\alpha\beta}\sim 1$.
In bidisperse suspensions the deviation from unity in $f_{\alpha\beta}$  is due to the size effects in HIs.

A simplification for the hydrodynamic interactions in bidisperse suspensions is to assume that the HIs are of a mean-field nature, and consequently the factors in Eq.~(\ref{eq:hqaa-scal}) and (\ref{eq:hqab-scal}) become $q$-independent, \ie,
\begin{align}
    f_\alpha(q;  \lambda, \phi, y) &\approx f_\alpha(\lambda,\phi,y) \label{eq:fa}\\
    f_{\alpha\beta}(q;  \lambda, \phi, y) &\approx f_{\alpha\beta}(\lambda,\phi,y).\label{eq:fab}
\end{align}
In this way, the monodisperse $\delta\gamma$ scheme is extended to bidisperse suspensions by introducing composition dependent scaling constants.
We call the resulting approximation scheme the rescaled $\delta\gamma$ scheme.
As we will see in \sectionname~\ref{sec:Results}, this simplification describes the SD measurement surprisingly well---providing an \textit{a posteriori} justification for Eq.~(\ref{eq:fa}) and (\ref{eq:fab}).
Note that the rescaling rules in Eq.~(\ref{eq:hqaa-scal}) and (\ref{eq:hqab-scal}) can be straightforwardly generalized to the polydisperse case with more than two different particle species.

\begin{figure}
 \includegraphics[width=.925\columnwidth]{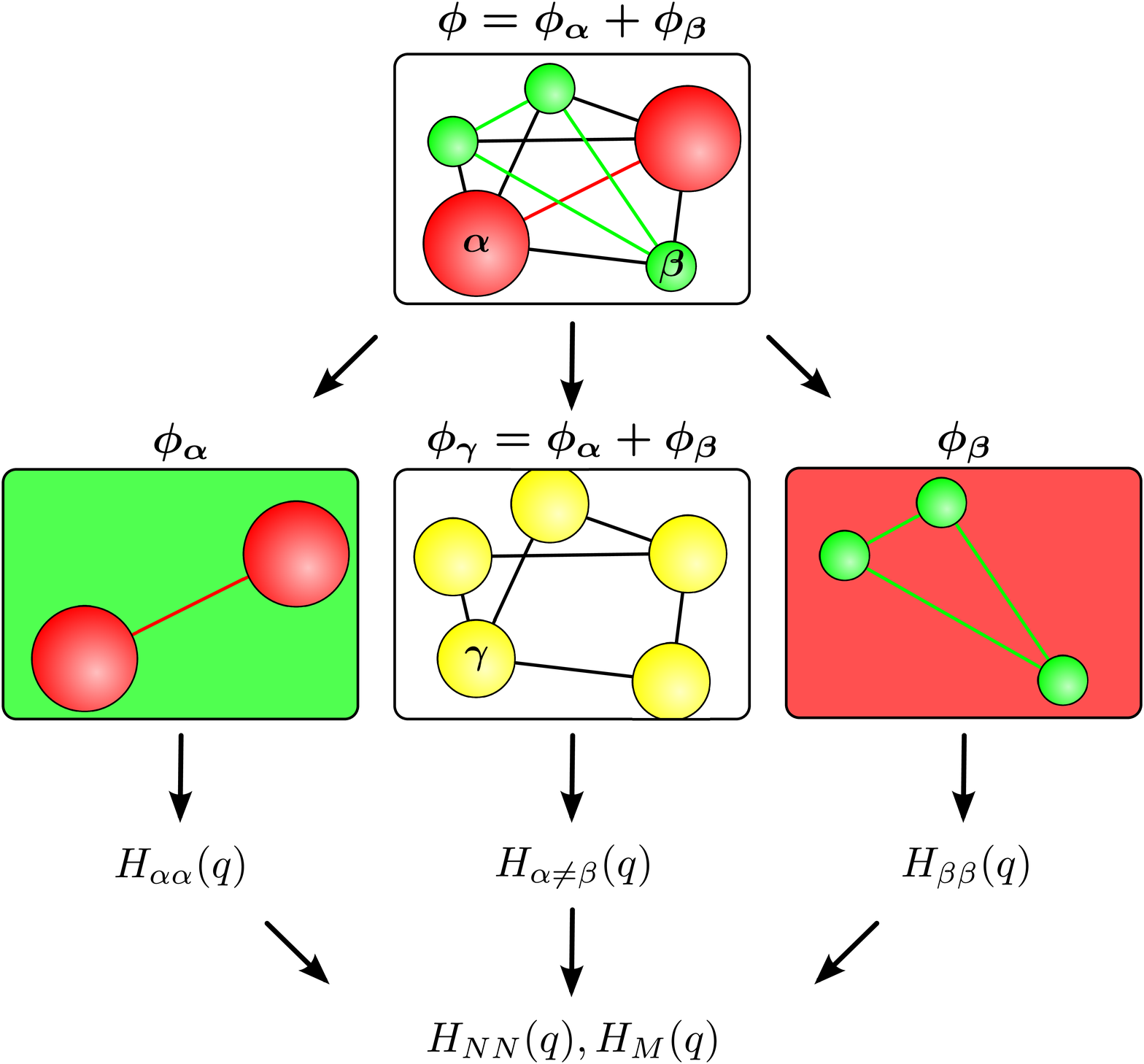} 
 \caption{(Color online) Schematic representation of the effective medium concept.
 Straight red, green and black curves indicate the $\alpha\alpha$, $\beta\beta$ and $\alpha \neq \beta$ correlations, respectively.
 Either species $\alpha, \beta$  is approximated as an effective structureless fluid for the other species to move in (left and right panels).
 The distinct species contributions ($\alpha \neq \beta$, central panel) are approximated by those of a hydrodynamically monodisperse fluid of
 fictitious $\gamma$-type particles in pure solvent. The size of $\gamma$-type particles is chosen such that $\phi_\gamma = \phi = \phi_\alpha + \phi_\beta$,
 and their center of mass positions coincide with those of the $\alpha-$ and $\beta-$ type particles in the bidisperse suspension (top panel).
 }
 \label{fig:schematic}
\end{figure}

Figure \ref{fig:schematic} succinctly illustrates the rescaled $\delta\gamma$ scheme. 
In computing the functions $H_{\alpha\alpha}(q)$, we ignore the particulate nature of species $\beta$ which is replaced by an effective medium for species $\alpha$ to move in (left and right panels in \figurename~\ref{fig:schematic}).
The effective translational free diffusion coefficient is therefore $f_\alpha d^\alpha_0$, and is expected to be smaller than the SSE diffusion coefficient $d_0^\alpha$ for diffusion in the pure solvent, leading to $f_\alpha < 1$.
The distinct species partial hydrodynamic function $H_{\alpha\beta}(q)$ for $\alpha \neq \beta$ is approximated by the corresponding function in a hydrodynamically monodisperse suspension of fictitious particles ($\gamma$-type particles in \figurename~\ref{fig:schematic}) in pure solvent, which occupy the same center of mass positions as the $\alpha$- and $\beta$- type particles in the bidisperse suspension. 
The size of the $\gamma$-type particles is chosen such that $\phi_\gamma = \phi = \phi_\alpha + \phi_\beta$.
We stress again that the fidelity of our approach cannot be easily estimated, but rather is validated \textit{a posteriori} by comparing with the SD simulation results.

For our rescaled $\delta\gamma$ scheme to be useful, estimations of the scaling factors $f_{\alpha}$ and $f_{\alpha\beta}$ are required.
To estimate the factor $f_{\alpha}$, recall that $f_{\alpha}d^\alpha_0$ describes the translational free diffusivity of one particle of species $\alpha$ in an effective medium of many $\beta$ particles.
Equivalently, for many $\alpha$ particles, $f_\alpha d_s(\phi_\alpha)/d_0$, where $d_s(\phi_\alpha)/d_0$ is the self-diffusivity of monodisperse suspensions at volume fraction $\phi_\alpha$, represents the species self-diffusivity $d^\alpha_s(\phi, \lambda, y)/d^\alpha_0$ in the bidisperse mixture, \ie,
\begin{equation}
\label{eq:est-fa}
  f_\alpha = \frac{d^\alpha_s(\phi, \lambda, y)/d^\alpha_0}{d_s(\phi_\alpha)/d_0},
\end{equation}
where the monodisperse self-diffusivity $d_s(\phi)/d_0$ is given in Eq.~(\ref{eq:ds_HS_monodisp_virial}), and the estimation of the species self-diffusivity is discussed next.
For the interspecies factor $f_{\alpha\beta}$, we assume the mean-field description of HIs is sufficient and the size effect is weak, \ie,
\begin{equation}
\label{eq:est-fab}
  f_{\alpha\beta} = 1.
\end{equation}
Note that both Eq.~(\ref{eq:est-fa}) and (\ref{eq:est-fab}) are physically motivated and are validated by the SD measurements in Section~\ref{sec:Results}.

The estimation of $f_\alpha$ in Eq.~(\ref{eq:est-fa}) requires an approximation of the species short-time self-diffusivity $d^\alpha_s/d^\alpha_0$ in the mixture. 
For dilute systems where HIs can be decomposed into sums of pairwise additive contributions, $d^\alpha_s/d^\alpha_0$ can be calculated to linear order in the volume fractions as~\cite{Batchelor1983, Zhang2002}
\begin{equation}\label{eq:ds_HS_mix_linvirial}
\frac{d_s^\alpha}{d_0^\alpha} = 1 + \sum_{\beta=1,2} I_{\alpha\beta} \phi_\beta + \mathcal{O}(\phi_1^2, \phi_2^2), 
\end{equation}
with the integrals
\begin{equation}\label{eq:ds_HS_mix_Integrals}
I_{\alpha\beta} = \dfrac{{(1 + \lambda_{\beta\alpha})}^3}{8 \lambda_{\beta\alpha}^3}
\int_2^\infty s^2 \left[ x_{11}^a(s) + 2 y_{11}^a(s) - 3 \right] ds
\end{equation}
in terms of $s = 2r/(a_\alpha + a_\beta)$ and $\lambda_{\beta\alpha} = a_\beta / a_\alpha$.  
The scalar hydrodynamic two-body mobility functions $x_{11}^a(s)$ and $2 y_{11}^a(s)$ describe the relative motions of two spheres in the direction parallel and orthogonal to a line that connects the sphere centers, respectively, and can be calculated with arbitrary precision \cite{Schmitz1982, Jeffrey1984, KimKarilla}.
A series expansion in the inverse particle separation yields the leading order far-field terms of the integrand
\begin{equation}\label{eq:twobodymob_asymptotic}
x_{11}^a + 2 y_{11}^a - 3 =
\dfrac{-60 \lambda_{\beta\alpha}^3}{{[s(1+\lambda_{\beta\alpha})]}^4} 
+\dfrac{480 \lambda_{\beta\alpha}^3 - 264 \lambda_{\beta\alpha}^5}{{[s(1+\lambda_{\beta\alpha})]}^6}
+\mathcal{O}(s^{-8}).
\end{equation}
Here, we employ the two-body mobility coefficients from ref.~\onlinecite{Jeffrey1984} up to $s^{-300}$ to ensure a smooth crossover to the analytically known close-contact (lubrication) expressions \cite{Jeffrey1982}.
For particle size-ratio $\lambda = 2$, numerical integration of \expressionname~\eqref{eq:ds_HS_mix_Integrals} yields the values
$I_{11} = I_{22} = -1.8315$, $I_{12} = -1.4491$ and $I_{21} = -2.0876$. 

Computation of the quadratic and higher order terms of the virial expansion in \expressionname~\eqref{eq:ds_HS_mix_linvirial} is an elaborate task,
even when three-body HIs are included in their leading-order far-field asymptotic form only \cite{Zhang2002}.
In place of such cumbersome computation of the ${d_s^\alpha}/{d_0^\alpha}$, we propose a simple Ansatz
\begin{equation}\label{eq:ds_HS_mix_adhoc}
\frac{d_s^\alpha}{d_0^\alpha} \approx 1 + \left(\sum_{\beta=1,2} I_{\alpha\beta} \phi_\beta\right) \times \left(1 + 0.1195\phi - 0.70\phi^2\right)   
\end{equation}
which reduces to the accurate expression in \expressionname~\eqref{eq:ds_HS_monodisp_virial} for $\lambda = 1$, and is correct to linear order in the volume fractions for all values of $\lambda$. 
In Eq.~(\ref{eq:ds_HS_mix_adhoc}), the effects of different particle sizes are incorporated in the linear term while the effects of different volume fractions are treated in a mean-field way, \ie, independent of the size ratio. 
It is important to note here that \expressionname~\eqref{eq:ds_HS_mix_adhoc} is merely an educated guess for the quadratic and cubic terms in the virial expansions of the ${d_s^\alpha}/{d_0^\alpha}$. 
The accuracy of \eqref{eq:ds_HS_monodisp_virial} will be tested by comparison to our SD results in \sectionname~\ref{sec:Results}.
With Eqs.~(\ref{eq:ds_HS_monodisp_virial}), (\ref{eq:est-fab}), and (\ref{eq:ds_HS_mix_adhoc}), the analytical estimation for $f_\alpha$ is
\begin{equation}
f_{\alpha} = \dfrac{\displaystyle{\rule[-7mm]{0mm}{6mm}1 + \left(\sum\limits_{\beta=1,2} I_{\alpha\beta} \phi_\beta\right) \times \left(1 + 0.1195\phi - 0.70\phi^2\right)}}
{\displaystyle{\rule[0mm]{0mm}{6mm}1 - 1.8315\phi_\alpha \left(1 + 0.1195\phi_\alpha - 0.70\phi_\alpha^2\right)}} \label{eq:f_alpha_beta}.
\end{equation}
%


\section{Results and discussions}\label{sec:Results}

In this section we compare results of the rescaled $\delta\gamma$ scheme described in \sectionname~\ref{sec:dg} to the results of the SD simulations outlined in \sectionname~\ref{sec:SD}.
For each  suspension composition, the SD simulations typically take a few days, while computations using the rescaled $\delta\gamma$ scheme only require at most a few minutes.
This great performance incentive renders the rescaled $\delta\gamma$ scheme more convenient for many applications.

The rescaled $\delta\gamma$ scheme relies on the monodisperse $\delta\gamma$ scheme to capture the structural features in the hydrodynamic functions of bidisperse suspensions, using bidisperse static structure factors as input.
The validity of this Ansatz can be directly validated by studying a bidisperse suspension where one of the species, say, species $\beta$, only influences the suspension structurally but not hydrodynamically, \ie,  $f_\alpha= 1$ in Eq.~(\ref{eq:hqaa-scal}).
An experimental realization of such system would be a mixture of hard-sphere particles and highly permeable porous but rigid particles of different size.
In the SD simulations, we generate a bidisperse suspension configuration and then exclude the inactive species $\beta$ from the hydrodynamic computations.
The resulting hydrodynamically monodisperse, but structurally bidisperse suspension's function $H(q)$ is influenced by the partial static structure factor $S_{\alpha\alpha}(q)$.

\begin{figure}[ht]
  \centering
  \includegraphics[width=3in]{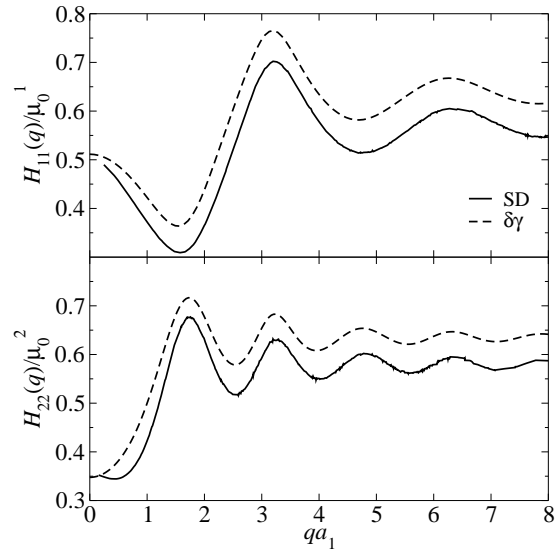}
  \caption{The partial hydrodynamic functions $H_{11}(q)$ and $H_{22}(q)$ for a bidisperse suspension of $\phi = 0.4$, $y = 0.5$, and $\lambda = 2$ with the respective other species being hydrodynamically inactive.
The hydrodynamic functions are scaled with the single particle mobility $\mu_{0}^{\alpha} = (6\pi\eta_0 a_{\alpha})^{-1}$ and the wave number is scaled with $a_1$, the radius of the smaller particles.}
  \label{fig:ghost-haa}
\end{figure}

Figure~\ref{fig:ghost-haa} compares the partial hydrodynamic functions $H_{\alpha\alpha}(q)$ of bidisperse suspensions containing hydrodynamically inactive particles from the rescaled $\delta\gamma $ scheme [Eq.~(\ref{eq:hqaa-scal}) with $f_{\alpha} = 1$] and the SD simulations. 
Recall that, for example, $H_{11}(q)$ corresponds to suspensions with  hydrodynamically inactive large particles.
Comparing to the SD measurements, the monodisperse $\delta\gamma$ scheme accurately captures the structural features in the hydrodynamic functions with structural input $S_{11}(q)$, including in particular the minimum in $H_{11}(q)$ for $qa_1 \approx 1.7$ due to cages formed by the large particles.
However, the monodisperse $\delta\gamma$ scheme systematically overestimates the magnitude of the hydrodynamic functions at all wave-numbers, since the species self-diffusivity in this case is different from the self-diffusivity in Eq.~(\ref{eq:ds_HS_monodisp_virial}) for monodisperse suspensions, due to the different suspension structures.

\begin{figure*}[ht]
  \centering
  \includegraphics[width=7in]{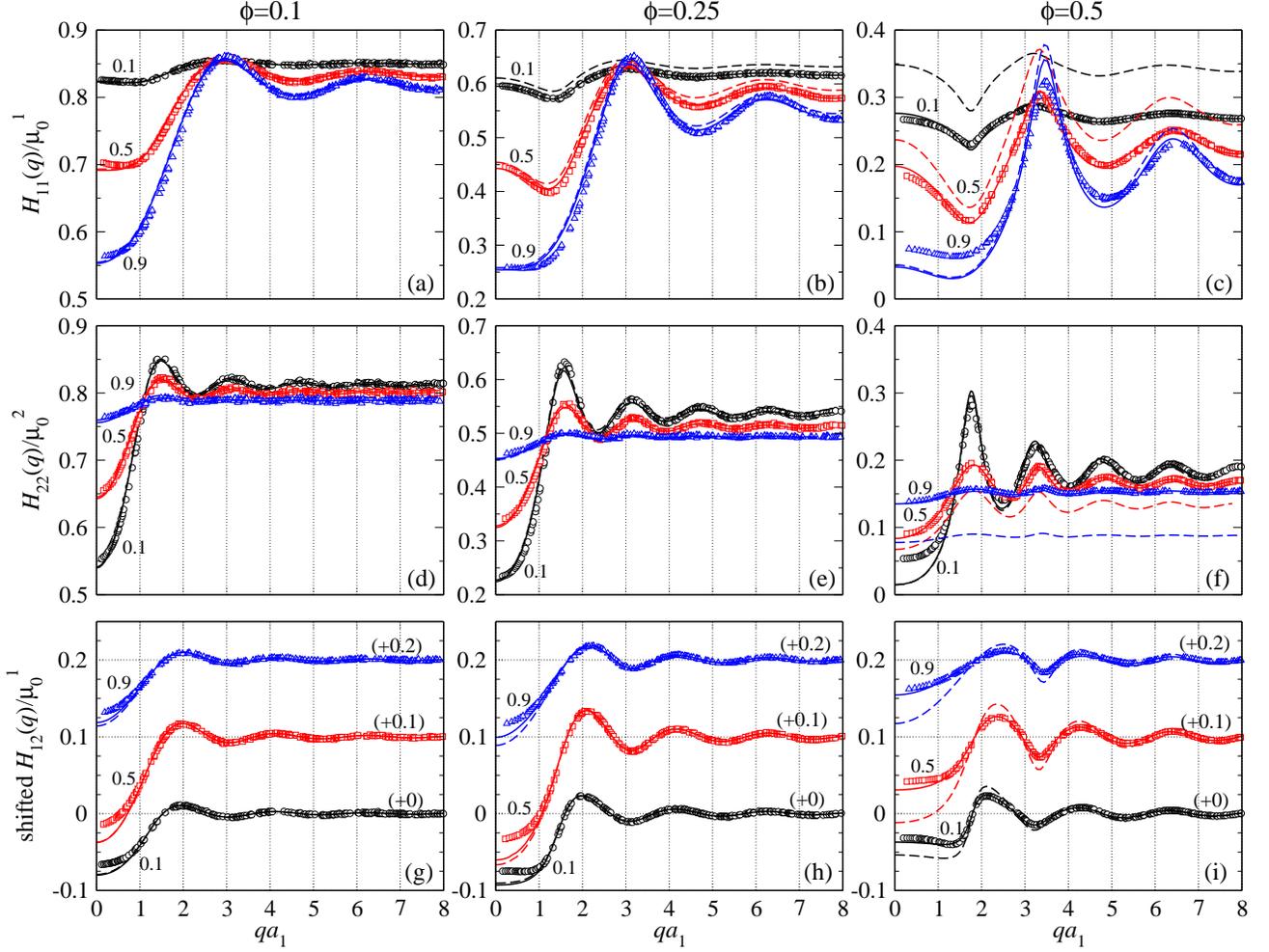}
  \caption{(Color online) The partial hydrodynamic functions $H_{\alpha\beta}(q)$ of bidisperse suspensions with full hydrodynamics.  The size ratio is $\lambda = 2$.
The top, middle, and bottom rows are $H_{11}(q)$ and $H_{22}(q)$, and $H_{12}(q)$, respectively.
The interspecies partial hydrodynamic functions $H_{12}(q)$ are shifted by $0.1$ for $y=0.5$ and by $0.2$ for $y=0.9$ for clarity (also indicated in the figure).
The left, middle, and right columns correspond to volume fractions $\phi = 0.1$, $0.25$, and $0.5$, respectively.
For each $\phi$ we show the SD measurements for composition $y=0.1$ ($\bigcirc$), $y=0.5$ ($\Box$), and $y=0.9$ ($\bigtriangleup$).  The results for the fitted $\delta\gamma$ scheme are shown as solid curves, and results of the parameter-free rescaled $\delta\gamma$ scheme with $f_\alpha$ from Eq.~(\ref{eq:f_alpha_beta}) and $f_{\alpha\beta}$ from Eq.~(\ref{eq:est-fab}) are shown as dashed curves.
}
  \label{fig:h112212}
\end{figure*}

Turning now to the true (structurally and hydrodynamically) bidisperse suspensions where both species are hydrodynamically active, \figurename~\ref{fig:h112212} features the SD measurements (symbols) of the partial hydrodynamic functions $H_{\alpha\beta}(q)$ for bidisperse suspensions with $\lambda = 2$ over a wide range of the compositions $y$ and total volume fractions $\phi$, covering both the dilute ($\phi=0.1$) and the concentrated ($\phi=0.5$) regimes.
The qualitative and quantitative aspects of the functions $H_{\alpha\beta}(q)$ are extensively examined and discussed in a companion paper~\cite{sd-bimodal_wang_2014}, and here we focus on the performance of the rescaled $\delta\gamma$ scheme.

\begin{figure*}[ht]
  \centering
    \includegraphics[width=7in]{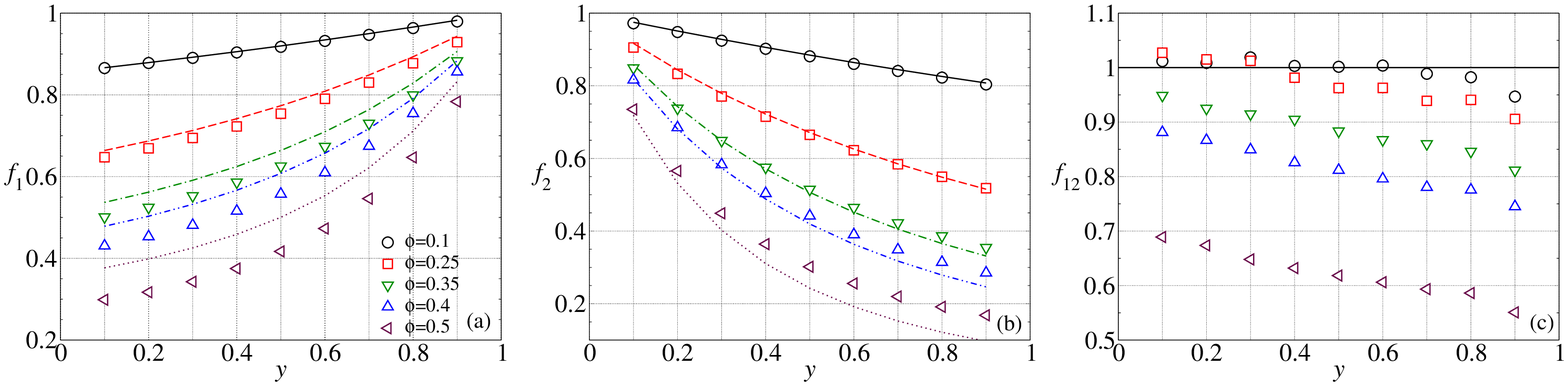}
  \caption{(Color online) The fitted $q$-independent scaling factors (a): $f_1$, (b): $f_2$, and (c): $f_{12}$ in the rescaled $\delta\gamma$ scheme for the bidisperse suspensions with $\lambda = 2$.  The curves are calculated according to \expressionname~(\ref{eq:f_alpha_beta}) for $f_\alpha$ with $\phi=0.1$ (solid), $0.25$ (dashed), $0.35$ (dash-dotted), $0.4$ (dash-double-dotted), and $0.5$ (dotted).
}
  \label{fig:f1f2f12}
\end{figure*}

We first discuss the central assumptions of the rescaled $\delta\gamma$ scheme: the wave-number independence of the fitting parameters $f_\alpha$ and $f_{\alpha\beta}$ in Eq.~(\ref{eq:fa}) and (\ref{eq:fab}), respectively.
The $q$-independent parameters $f_\alpha$ and $f_{\alpha\beta}$ were computed by least-square fitting the SD measurements and the rescaled $\delta\gamma$ scheme as in Eq.~(\ref{eq:hqaa-scal}) and (\ref{eq:hqab-scal}).
The fitted partial hydrodynamic functions are presented as solid curves in \figurename~\ref{fig:h112212}.
For $H_{\alpha\alpha}(q)$, the fitted data capture all the qualitative and most quantitative features in the SD measurements at all $q$ for both species.
The best agreement is found at $y=0.5$, where both species are presented in large amount for the mean-field description of the HIs to be valid.
For more asymmetric compositions, such as at $y=0.1$ and $y=0.9$, the agreement deteriorates slightly at low $q$ with increasing $\phi$.
For the dilute suspensions at $\phi=0.1$, we find excellent agreement between the fitted functions and the SD measurements.
At $\phi=0.25$, despite the excellent overall agreement for both species, the discrepancies are slightly more pronounced for the smaller species.
The mean-field description is more appropriate for the hydrodynamic environment of the large particles, as each of them is surrounded by multiple small particles.
On the other hand, the small particles are strongly affected by the presence of large particles, and the respective hydrodynamic environment exhibits more fluctuations.
This leads to the slight disagreement in $H_{11}(q)$ at $y=0.9$ in figure~\ref{fig:h112212}(b).
At $\phi=0.5$, the accuracy of the $\delta\gamma$ scheme breaks down since the unaccounted hydrodynamic scattering diagrams become important.
However,  despite some disagreements the fitted scheme still captures many qualitative features of $H_{\alpha\alpha}(q)$.
The disagreements are particularly apparent in the low $q$ limit with asymmetric compositions, \eg, $H_{11}(q)$ at $y=0.9$ in figure~\ref{fig:h112212}(c) and $H_{22}(q)$ at $y = 0.1$ in figure \ref{fig:h112212}(f).
In these cases, the $q$-independent scaling factor $f_{\alpha}$ is not sufficient to describe the hydrodynamic interactions from the minority species $\beta$. 
For $H_{\alpha\beta}(q)$ ($\alpha\neq \beta$) shown in figure \ref{fig:h112212}(g)--(i),  the agreement between the measured and fitted $H_{12}(q)$ is excellent for all $\phi$ except at small $q$.
Note that the modulations of $H_{12}(q)$ first increase from $\phi=0.1$ to $\phi=0.25$ due to the enhancement of hydrodynamic interactions, and then decrease from $\phi = 0.25$ to $\phi = 0.5$, possibly due to hydrodynamic shielding effects.
The overall $q$-variations in $H_{12}(q)$ are small compared to $H_{11}(q)$ and $H_{22}(q)$.
Overall, the agreement between the SD measurement and the fitted scheme validates the assumption of $q$-independence of $f_\alpha$ and $f_{\alpha\beta}$,
up to relatively high volume fractions.

The fitted $q$-independent scaling factors $f_1$, $f_2$, and $f_{12}$ as a function of the composition $y$ for bidisperse suspensions with $\lambda = 2$ at different volume fractions $\phi$ are presented in \figurename~\ref{fig:f1f2f12}.
As expected, at a fixed volume fraction $\phi$, $f_\alpha$ decreases monotonically from $1$ with the increasing presence of the other species $\beta$.
At a fixed value of $y$, $f_\alpha$ also decreases from $1$ when the volume fraction $\phi$ is increased.  
Both decreasing trends in $f_\alpha$ are due to the enhanced HIs from the other species.
The scaling factor $f_{12}$ for the interspecies hydrodynamic interactions exhibits more peculiar behaviors.  
For $\phi = 0.1$ and $0.25$, the factor $f_{12}$ is close to unity, suggesting that the mean-field hydrodynamic interaction assumption in the rescaled $\delta\gamma$-scheme is valid.
However, $f_{12}$ does become smaller with increasing $y$, \ie, for $H_{12}(q)$, adding larger particles to the suspension is not equivalent to adding smaller particles,
which becomes particularly clear for $\phi \geq 0.25$ in \figurename~\ref{fig:f1f2f12}(c).
For $\phi = 0.4$ and $0.5$, $f_{12}$ becomes much smaller than unity and decreases monotonically with increasing $y$.  
At these volume fractions, it appears that $f_{12}$ is extremely sensitive to the presence of the other species in the mixture, as we expect $f_{12}$ to recover to unity when $y\rightarrow 0$ or $y\rightarrow 1$.

The $f_1$ and $f_2$ predicted by \expressionname~(\ref{eq:f_alpha_beta}) are shown in \figurename~\ref{fig:f1f2f12}(a) and (b) as curves.  
The predicted $f_1$ agrees well with the fitted value up to $\phi = 0.25$, and at higher volume fractions, the equation overestimates $f_1$ by $10\%$ at $\phi=0.35$ and $y=0.1$ and by $20\%$ at $\phi=0.45$ and $y=0.1$.
The predicted  $f_2$ for the larger species, however, agrees well with the fitted value up to $\phi=0.4$ at all compositions except when $y$ is close to unity.
Since Eq.~(\ref{eq:f_alpha_beta}) is motivated by a mean-field model of $d^\alpha_s/d^\alpha_0$, \expressionname~(\ref{eq:ds_HS_mix_adhoc}), \figurename~\ref{fig:f1f2f12} again suggests that the larger particles in bidisperse suspensions experience the mean field from the small particles, while the hydrodynamic environment of the smaller particles shows stronger fluctuations.
For practical purposes, from figure \ref{fig:f1f2f12} we note that the parameter-free analytical estimation of $f_{\alpha}$ and $f_{\alpha\beta}$
is satisfactory up to $\phi\sim 0.35$--$0.4$ at all compositions, for $\lambda = 2$.

The parameter-free partial hydrodynamic functions, predicted by the rescaled $\delta\gamma$ scheme with factors $f_\alpha$
from Eq.~(\ref{eq:f_alpha_beta}) and $f_{12}$ from Eq.~(\ref{eq:est-fab}), are presented in figure~\ref{fig:h112212} as dashed curves.
The agreement with the SD measurements is satisfactory for $H_{\alpha\beta}(q)$ at all compositions at $\phi=0.1$ and $0.25$.  
In figure~\ref{fig:h112212}(b) the predicted $f_1$ slightly overestimates $H_{11}(q)$ at $y=0.1$ at $\phi=0.25$, primarily due to the overestimation of the small particle diffusivity in Eq.~(\ref{eq:ds_HS_mix_adhoc}).
At $\phi=0.5$,  the prediction breaks down, and the discrepancy is most pronounced at $y=0.1$ for the overestimation of $H_{11}(q)$ in figure~\ref{fig:h112212}(c) and at $y=0.9$ for the underestimation of $H_{22}(q)$ in figure~\ref{fig:h112212}(f). 
Moreover, Eq.~(\ref{eq:est-fab}) overestimates the $q$-modulations in $H_{12}(q)$ in all compositions at $\phi = 0.5$ in figure~\ref{fig:h112212}(i), as the hydrodynamic shielding in dense systems cannot be captured by $f_{12}=1$.

\begin{figure*}[ht]
  \centering
  \includegraphics[width=7in]{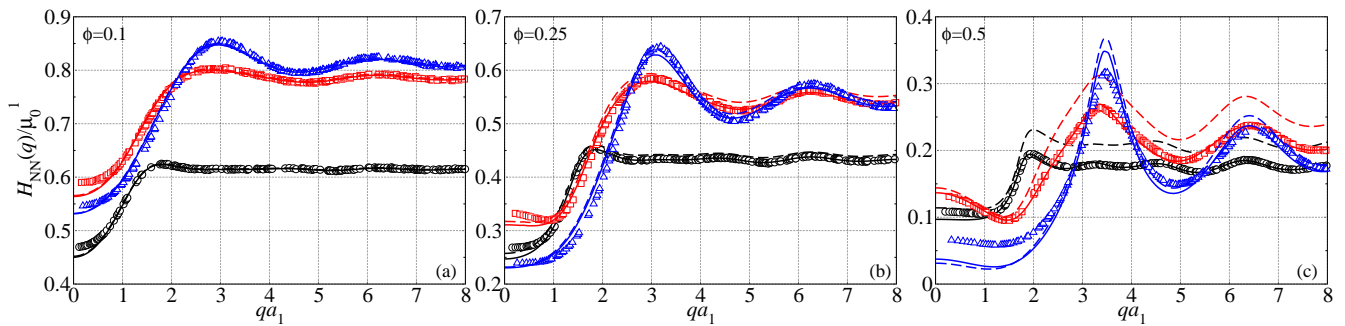}
  \caption{(Color online) The number-number hydrodynamic functions $H_{NN}(q)$ for bidisperse suspensions with $\lambda = 2$ and full hydrodynamics for volume fractions (a): $\phi = 0.1$, (b):$\phi=0.25$, and (c): $\phi=0.5$.
For each $\phi$, we show the SD measurements for composition $y=0.1$ ($\bigcirc$), $y=0.5$ ($\Box$), and $y=0.9$ ($\bigtriangleup$). The $H_{NN}(q)$ from the $\delta\gamma$ scheme with fitted $f_\alpha$ and $f_{\alpha\beta}$ are shown as solid curves, and the results of the parameter-free theory with $f_\alpha$ according to \expressionname~(\ref{eq:f_alpha_beta}) and $f_{\alpha\beta}$ according to \expressionname~(\ref{eq:est-fab}) are shown as dashed curves.
}
  \label{fig:hnn}
\end{figure*}

In practice, individual partial hydrodynamic functions $H_{\alpha\beta}(q)$ cannot be conveniently measured in scattering experiments and the measured quantity $H_M(q)$ is a weighted average of the $H_{\alpha\beta}(q)$. Note from \expressionsname~\eqref{eq:H_NN} and \eqref{eq:H_M}, that $H_M(q)$ differs from the similar number-number hydrodynamic function $H_{NN}(q)$
only trough its dependence on the particle-specific scattering amplitudes $f_\alpha(q)$.
To test the accuracy of the rescaled $\delta\gamma$ scheme, it is sufficient to test its predictions of $H_{NN}(q)$.
In \figurename~\ref{fig:hnn} we compare the $H_{NN}(q)$ from the SD measurements and from the rescaled $\delta\gamma$ scheme, with factors $f_\alpha$ and $f_{\alpha\beta}$ obtained from optimal least square fittings (solid curves) and from the parameter-free analytic \expressionsname~(\ref{eq:f_alpha_beta}) and (\ref{eq:est-fab}) (dashed curves).
Results for the same bidisperse suspensions are depicted in \figuresname~\ref{fig:hnn} and \ref{fig:h112212}.
For $\phi=0.1$, the rescaled $\delta\gamma$ scheme captures the SD results with high precision in the entire $q$-range, at all studied compositions $y$.
Small discrepancies occur most noticeably in the $q\rightarrow 0$ limit.
At $\phi = 0.25$, the difference in $H_{NN}(q)$ from both the fitted and the parameter-free analytical expression
is less than $5\%$ in the entire $q$-range, which demonstrates the validity of our proposed rescaling rules for the $\delta\gamma$ scheme.
For the very dense suspensions, $\phi=0.5$, we see how the rescaled $\delta\gamma$ scheme breaks down.  
With the fitted $f_{\alpha}$ and $f_{\alpha\beta}$, the scheme is only capable of capturing the qualitative features in the measured $H_{NN}(q)$.
With the $f_{\alpha}$ and $f_{\alpha\beta}$ from Eq.~(\ref{eq:f_alpha_beta}) and (\ref{eq:est-fab}), the scheme exhibits significant differences from the SD measurements with decreasing $y$.

For larger size ratios, our tests on $\lambda = 4$ shows the diffusivity approximation of Eq.~(\ref{eq:est-fa}) remains valid but the approximate expression [Eq.(\ref{eq:f_alpha_beta})] breaks down at $\phi = 0.25$ and $y=0.5$.  This is because Eq.~(\ref{eq:ds_HS_mix_adhoc}) is insufficient to describe the species short-time self-diffusivity $d^\alpha_s/d^\alpha_0$.  It appears that an expression to estimate the species diffusivity in dense suspensions with disparate size ratios is the key to the success of the rescaled $\delta\gamma$ scheme.

\section{Conclusions}\label{sec:Conclus}

In this work we have proposed a rescaled $\delta\gamma$ scheme to compute approximations of the partial hydrodynamic functions $H_{\alpha\beta}(q)$ in colloidal mixtures.
We found that the $H_{\alpha\beta}(q)$ from the Stokesian Dynamics measurements differs from the $\delta\gamma$ scheme with appropriate structural input by a $q$-independent factor, suggesting that the hydrodynamic environment for one species can be described as a mean field due to the HIs from the other species and the solvent.
This constitutes the fundamental assumption of the rescaled $\delta\gamma$ scheme.

We extensively tested the rescaled $\delta\gamma$ scheme with the SD simulation measurements for bidisperse suspensions over a wide range of volume fractions $\phi$ and compositions $y$, and provided approximate analytical estimates for the scaling factors $f_\alpha$, and $f_{\alpha\beta}$.
Comparing with the SD measurements, the rescaled $\delta\gamma$ scheme with analytical scaling factors can accurately predict the number-number hydrodynamic function $H_{NN}(q)$ up to $\phi \approx 0.4$ at all studied composition ratios $y$, for a particle-size ratio as high as $\lambda = 2$.

The proposed rescaled $\delta\gamma$ scheme is the first semi-analytical method for estimating the bidisperse hydrodynamic functions up to $\phi=0.4$,
and it can be readily extended to polydisperse and charged systems. It will be a valuable tool for interpreting dynamic scattering experiments of moderately dense bidisperse systems.

\begin{acknowledgments}
We thank Karol Makuch for his helpful comments and discussions of the $\delta\gamma$ scheme. 
M.W. acknowledges support by a Postgraduate Scholarship (PGS) of the Natural Sciences and Engineering Research Council of Canada (NSERC), and the National Science Foundation (NSF) grant CBET-1337097. 
M.H. acknowledges support by a fellowship within the Postdoc-Program of the German Academic Exchange Service (DAAD).
\end{acknowledgments}

\bibliographystyle{unsrt}
\bibliography{HS_mix}

\end{document}